\documentclass[12pt]{spieman}  % 12pt font required by SPIE;
\usepackage{amsmath,amsfonts,amssymb}
\usepackage{graphicx}
\usepackage{setspace}
\usepackage{tocloft}
\usepackage{varwidth}

\usepackage{aas_macros}

\usepackage{url}

\title{XRISM Pre-Pipeline and Singularity: Container-Based Data Processing for the X-Ray Imaging and Spectroscopy Mission and High-Performance Computing}

\author[a, *]{Satoshi Eguchi}
\author[b, c]{Makoto Tashiro}
\author[b, c]{Yukikatsu Terada}
\author[d]{Hiromitsu Takahashi}
\author[e]{Masayoshi Nobukawa}
\author[c]{Ken Ebisawa}
\author[c]{Katsuhiro Hayashi}
\author[c]{Tessei Yoshida}
\author[c]{Yoshiaki Kanemaru}
\author[c]{Shoji Ogawa}
\author[f]{Matthew P. Holland}
\author[g, f, h]{Michael Loewenstein}
\author[i]{Eric D. Miller}
\author[j, f, h]{Tahir Yaqoob}
\author[k, f]{Robert S. Hill}
\author[k, f]{Morgan D. Waddy}
\author[k, f]{Mark M. Mekosh}
\author[k, f]{Joseph B. Fox}
\author[k, f]{Isabella S. Brewer}
\author[k, f]{Emily Aldoretta}
\author[l]{Yuusuke Uchida}
\author[c]{Nagomi Uchida}
\author[c]{Kotaro Fukushima}

\affil[a]{Kumamoto Gakuen University, Faculty of Economics, Department of Economics, 2-5-1 Oe, Chuo-ku, Kumamoto, Japan, 862-8680}
\affil[b]{Saitama University, Graduate School of Science and Engineering, Shimo-Okubo 255, Sakura-ku, Saitama, Japan, 338-8570}
\affil[c]{Institute of Space and Astronautical Science, Japan Aerospace Exploration Agency, 3-1-1 Yoshinodai, Chuo-ku, Sagamihara, Japan, 252-5210}
\affil[d]{Hiroshima University, School of Science, 1-3-1 Kagamiyama, Higashi-Hiroshima, Japan, 739-8511}
\affil[e]{Nara University of Education, Department of Teacher Training and School Education, Takabatake-cho, Nara, Japan, 630-8528}
\affil[f]{National Aeronautics and Space Administration, Goddard Space Flight Center, 8800, Greenbelt, Maryland, United States, 20771}
\affil[g]{University of Maryland, College Park, Maryland, United States, 20742}
\affil[h]{Center for Research and Exploration in Space Science and Technology (CRESST), Greenbelt, Maryland, United States, 20771}
\affil[i]{Massachusetts Institute of Technology, MIT Kavli Institute for Astrophysics and Space Research, 70 Vassar St, Cambridge, United States, 02139}
\affil[j]{University of Maryland, Baltimore County, Baltimore, Maryland, United States, 21250}
\affil[k]{ADNET Systems, Inc., 6720B Rockledge Drive, Suite \#504, Bethesda, Maryland, United States, 20817}
\affil[l]{Tokyo University of Science, Faculty of Science and Technology, 2641 Yamazaki, Noda-shi, Chiba, Japan, 278-8510}

\cftpagenumbersoff{figure}
\cftpagenumbersoff{table}
\begin{document}
\maketitle

\begin{abstract}
The X-Ray Imaging and Spectroscopy Mission (XRISM) is the seventh Japanese X-ray observatory
whose development and operation are in collaboration with universities and research institutes in Japan,
the United States, and Europe, including JAXA, NASA, and ESA.
The telemetry data downlinked from the satellite are reduced to scientific products
using pre-pipeline (PPL) and pipeline (PL) software running on standard Linux virtual machines (VMs)
for the JAXA and NASA sides, respectively.
OBSIDs identified the observations, and we had 80 and 161 OBSIDs to be reprocessed
at the end of the commissioning period and performance verification and calibration period, respectively.
The combination of the containerized PPL utilizing Singularity of a container platform
running on the JAXA's ``TOKI-RURI'' high-performance computing (HPC) system and working disk images
formatted to ext3 accomplished a $33 \times$ speedup in PPL tasks over our regular VM.
Herein, we briefly describe the data processing in XRISM and our porting strategies for PPL
in the HPC environment.
\end{abstract}

% Include a list of up to six keywords after the abstract
%%\keywords{containers, high-performance computing (HPC), astroinformatics, X-ray, pipelines}

% Include email contact information for the corresponding author
{\noindent \footnotesize\textbf{*}Satoshi Eguchi  \linkable{sa-eguchi@kumagaku.ac.jp} }

%%\begin{spacing}{2}   % use double spacing for rest of manuscript
\begin{spacing}{1}   % use double spacing for rest of manuscript

\section{Introduction} \label{sect:intro}

The X-Ray Imaging and Spectroscopy Mission (XRISM) is the seventh Japanese X-ray observatory
and the successor of the Hitomi X-ray satellite.
It was launched at 8:42 a.m. on September 7, 2023, in Japan Standard Time with the H-IIA Launch
Vehicle No.~47 from Tanegashima Space Center.
The project is led by the Japan Aerospace Exploration Agency (JAXA) and the National
Aeronautics and Space Administration (NASA) in collaboration with the European Space
Agency (ESA) and other partners.
XRISM carries the revolutionary X-ray microcalorimeter array with a high-energy resolution
($\simeq 5 \ \text{eV}$; Resolve) and an X-ray CCD camera with a large field of view
($38 \times 38 \ \text{arcmin}^{2}$; Xtend).
The first light observations of supernova remnant N132D in the Large Magellanic Cloud\cite{Audard2024}
and the active galactic nucleus NGC 4151\cite{xrism_ngc4151} with Resolve,
and the galaxy cluster Abell 2319\cite{Mori2024} with Xtend promise that XRISM provides a leap
in our understanding of the formation and evolution of the Universe, galaxies, compact objects,
and supernovae.

The mission schedule is as follows: the critical operation and ``initial phase commissioning period''
(the ``commissioning period'' hereafter; lasting three months after the launch),
the ``nominal phase initial calibration and performance verification period''
(the ``PV period'' hereafter; lasting seven months after the commissioning period),
and the ``nominal phase nominal observation period''
(the ``nominal observation period'' hereafter; lasting 26 months after the PV period).
The nominal observation period will conclude in September 2026,
after which XRISM will undergo a review to evaluate whether it can transition into the late phase,
that is, the extended observation phase.
As of January 2025, XRISM is in the general observer program cycle 1 (GO-1) during the nominal observation period.

The Science Operations Center (SOC) at JAXA, Science Data Center (SDC) at NASA,
and European Space Astronomy Centre (ESAC) at ESA share the science operation tasks
in XRISM, and the first two centers are responsible for data processing and distribution\cite{Terada2021}.
XRISM adopts a ``two-stage'' data reduction: pre-pipeline (PPL) process in SOC and subsequent
pipeline (PL) process in SDC.
Although the software was designed carefully prior to the launch,
further improvements are still underway;
downlinked observation data are regularly processed using the latest versions of PPL and PL
at that point.
When a significant update to either PPL or PL occurs at some point during a mission period
(this has happened a few times), the quality of the products distributed to scientists
changes from that point on.
As PPL processing cannot be applied to end users,
all data should be reprocessed using the latest software at certain points,
such as at the end of the commissioning and PV periods of the mission
(approximately once a year).

On the other hand, high-performance computing (HPC) is required to complete these reprocessing tasks
within a realistic time (less than one week),
which requires additional work to port the software to the HPC system.
To this end, we developed new and efficient porting methods that utilize Singularity\cite{Kurtzer2017},
which is a container platform.
Containers are lightweight virtualization technologies (VTs) compared to virtual machines (VMs).
Packaging our ordinal PPL environment with core system libraries, such as the GNU C Library (glibc), into a Singularity disk image allows
the PPL software to run on the JAXA ``TOKI-RURI'' HPC system\cite{JSS3} with minimal modifications.
Herein, we elaborate on our methods and processing results using the HPC system.

The remainder of this paper is organized as follows.
Section~\ref{sect:ppl-pl} briefly describes regular data processing using PPL and PL.
Section~\ref{sect:hpc-ppl-backgrounds} summarizes VTs in contrast
to VMs and containers, Singularity, and our motivations and methods.
Section~\ref{sect:results} discusses the processing results of TOKI-RURI.
Section~\ref{sect:summary} summarizes this study.

\section{Ordinary Data Processing for Pipeline Products} \label{sect:ppl-pl}

\begin{figure}
 \centering
 \includegraphics[clip,keepaspectratio,width=0.95\linewidth]{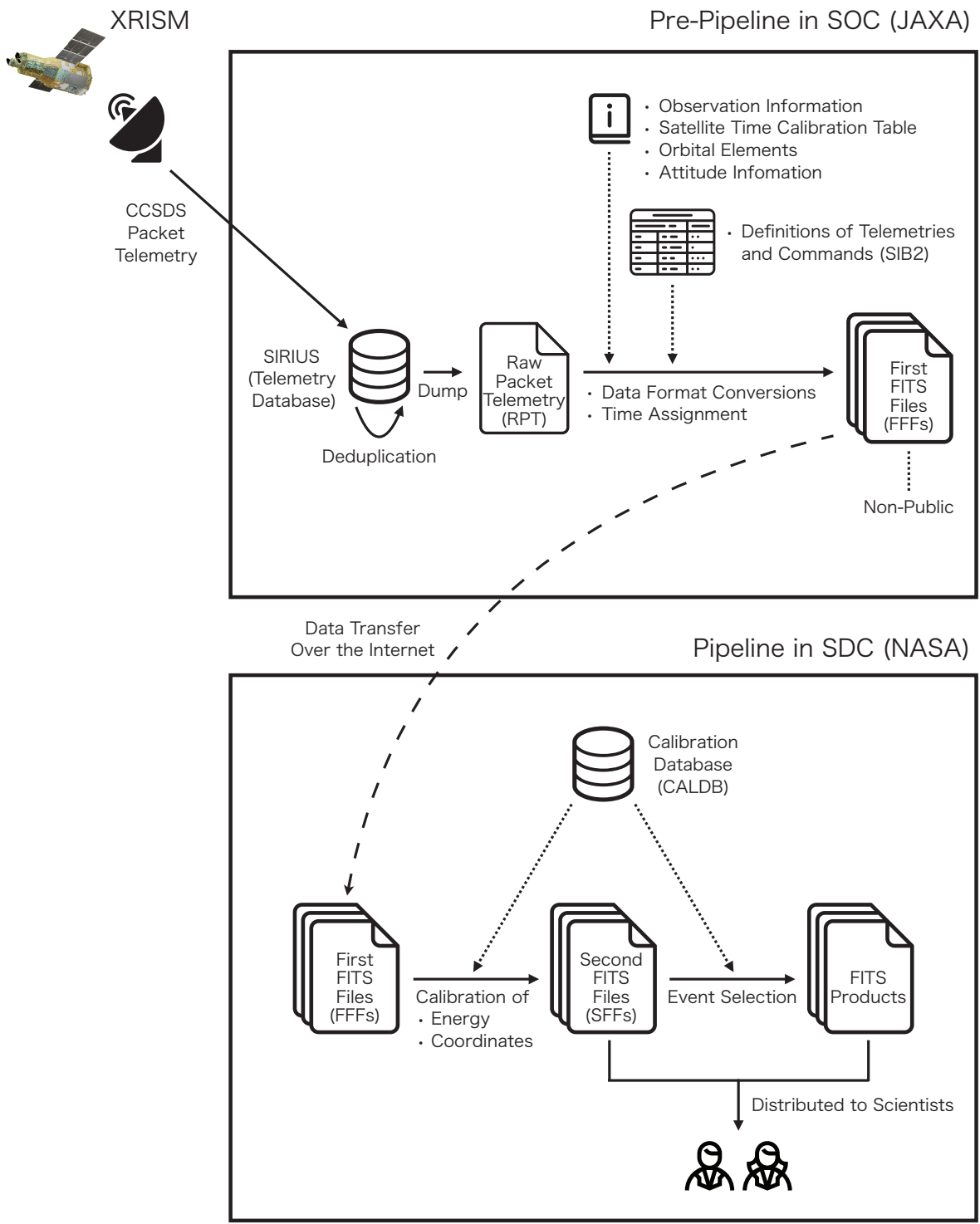}
 \caption{Schematic of data processing in XRISM.
  Telemetry data downlinked from XRISM are converted into FITS products with the pre-pipeline
  and pipeline software in SOC and SDC, respectively.
  Both centers distribute a set of SFFs and FITS products to the principal investigator
  of the observation.
  \label{fig:data-processing-diagram}}
\end{figure}

Figure~\ref{fig:data-processing-diagram} shows a schematic of the data reduction for
pipeline products in XRISM (see Refs.~\citenum{Terada2021} and \citenum{Eguchi2022} for details).
Data accumulated from payload instruments, including X-ray detectors,
in the form of Spacecraft Monitor and Control Protocol (SMCP)\cite{Yamada2008} messages
into an onboard data recorder (DR) are downlinked to the tracking and control stations
on the ground in the space packet format defined by the Consultative Committee
for Space Data Systems (CCSDS).
These packets are stored in a Scientific Information Retrieval and Integrated Utilization System
(SIRIUS)\cite{Okada2012}, which is a telemetry database operated by the Science Satellite Operation and Data
Archive Unit (C-SODA) in JAXA, along with information on communication antennae.
In regular data processing, we use the ``merged mode'' for SIRIUS, where SIRIUS ignores the antenna information
and behaves as if all packets in the system had been downlinked with one virtual antenna.
In this mode, SIRIUS dedupes the packets;
the same memory addresses in the DR can be read multiple times during daily operations for robustness,
and this functionality removes these duplicate packets.
An explicit specification of an antenna to SIRIUS allows us to access the removed packets.

During the first stage of the PPL software managed by SOC, which runs on a standard Linux
(RedHat Enterprise Linux 7) VM named ``Reformatter'' featuring 4-core virtual central processing units (CPUs)
and 64 GB of virtual memory hosted on a physical server with dual Intel Xeon E5-2698 v4 processors and 256 GB of memory in C-SODA,
the CCSDS packets corresponding to a single observation identified by OBSID---a nine-digit unique integer
where the highest digit indicates the observation type, such as commissioning (\texttt{=0}) and calibration (\texttt{=1}) observations---are dumped
into a single Flexible Image Transport System (FITS)\cite{Wells1981, Pence2010}
file with variable-length arrays called ``Raw Packet Telemetry'' (RPT).
In the latter stage of PPL, the CCSDS packets in the RPT are reconstructed into SMCP messages
and are then compiled into essential raw values in stages by referencing the definition files for
telemetry and commands (SIB2\cite{Yamada2008,Matsuzaki2012}; confidential) and satellite and observation information.
These raw values are then reduced to multiple ``First FITS Files'' (FFFs).
The typical PPL processing time is $\simeq 3$ hours/OBSID.
The FFFs are transferred to SDC.

The PL software in SDC calibrates the raw values in the FFFs into physical quantities
such as the coordinates and pulse height invariants in stages.
Referencing the calibration database, PL creates ``Second FITS Files'' (SFFs).
PL also generates ready-for-analysis FITS products (cleaned-event FITS files) by filtering out
low-quality events in the SFFs and applying good-time intervals.
A set of SFFs and FITS products is distributed to the principal investigator of the observation,
whereas the RPT and FFFs as files are hidden inside the XRISM project;
all columns in the FFFs are transferred to the SFFs, and users can access all the information
in the FFFs through the corresponding SFFs.
In other words, the only difference between FFFs and SFFs is
whether the correct values are entered into the FITS tables;
they are of the same size, except for the \texttt{HISTORY} and \texttt{COMMENT} keywords in their respective FITS headers.
Including intermediate information, no data are lost during PPL and PL processing.
From the perspective of the principal investigator as an end user,
only a few columns ($\sim 100$ MiB in size) in the FITS products (a few tens of GiB) are  necessary for analysis.

\section{Motivations and Strategies for the HPC version of PPL} \label{sect:hpc-ppl-backgrounds}

After the launch and the subsequent short critical operation period, XRISM experienced two essential
milestones: the commissioning and PV periods.
PPL and PL software improved aggressively in the first two periods
as it was fundamental for the instrument teams to examine whether their instruments were functioning properly.
This also implies that all observational data must have been simultaneously reprocessed
with the latest versions of PPL and PL at the end of each period to homogenize the quality of
the PPL and PL products in preparation for scientific publication.
Concurrently, there were 80 and 161 OBSIDs at the end of the commissioning
and PV periods, respectively, and they were unfeasible tasks for our ordinary VMs
(a temporary change to the VM settings according to the PPL tasks was not accommodated by C-SODA).
This has strongly motivated SOC to explore a method to boost PPL tasks in an HPC system
such as a supercomputer.
On the other hand, given that SIB2 is classified information and XRISM's operation is scheduled to continue
until September 2026, as of January 2025, developing a rapid method that could be implemented using resources
available in JAXA was deemed essential.
RedHat Enterprise Linux 7 or an alternative was also required for PPL because some programs can only be compiled with
and tested using the GNU Compiler Collection (GCC) 4.8.
We required parallelization in a unit of OBSID and did not intend to implement parallelizable
algorithms that use Open Multi-Processing and Message Passing Interface (MPI), among others, in PPL.

\subsection{JAXA Supercomputer System Generation 3} \label{sect:jss3}

The JAXA Supercomputer System Generation 3 (JSS3)\cite{JSS3}, which began operating in 2020,
is the infrastructure for numerical simulations and large-scale data analyses.
It consists of
\begin{itemize}
 \item TOKI-SORA: the leading computing platform equipping A64FX\cite{A64FX}-based CPUs
       with the theoretical peak performance of 19.4 PFLOPS,
 \item TOKI-RURI and TOKI-TRURI: x86-64 architecture clusters running Rocky Linux 8 for general-purpose computing,
       including machine learning, with the theoretical peak performance of 1.24 PFLOPS and 145 TFLOPS, respectively,
 \item TOKI-FS: a storage system with 10-PB non-volatile memory express (NVMe) solid-state drives (SSDs)
       and 40-PB hard disks (HDDs) in total, powered by the Lustre distributed parallel file system\cite{Lustre} (FS),
 \item J-SPACE: the archiving infrastructure.
\end{itemize}
There is also a 0.4-PB storage system called TOKI-TFS,
which connects to TOKI-TRURI.
For TOKI-SORA and TOKI-RURI, TOKI-FS serves as the sole hot storage;
all input and output files must reside in TOKI-FS.
It should be noted that these systems are completely isolated from the Internet.
To use them, one should log in to one of the login nodes through a virtual private network.
Furthermore, only SSH communication from the Internet to the login node is permitted.
The firewall prohibits communication in the reverse direction
(see the bottom half of Fig.~\ref{fig:network-overview}).

\begin{table}
 \caption{Specifications of the TOKI-RURI system} \label{tab:toki-ruri}
 \centering
 \begin{tabular}{|c||c|c|c|c|}
  \hline
  Type & TOKI-XM & TOKI-LM & TOKI-ST & TOKI-GP \\
  \hline
  \hline
  Number of Nodes & 2 & 7 & 375 & 32 \\
  \hline
  CPU (Intel Xeon) & Gold 6240L ($\times 2$) & \multicolumn{3}{c|}{Gold 6240 ($\times 2$)} \\
  \hline
  Memory & 6 TiB & 1.5 TiB & 192 GiB & 384 GiB \\
  \hline
  GPGPU (NVIDIA) & \multicolumn{3}{c|}{Quadro P4000 ($\times 1$)} & Tesla V100 SXM2 ($\times 4$) \\
  \hline
 \end{tabular}
\end{table}

TOKI-RURI is the best choice, as each node can be considered a standard personal computer (PC)
running Linux.
TOKI-RURI is further divided into four types as summarized in Table~\ref{tab:toki-ruri}.
Thus, the TOKI-RURI system has 14,976 physical CPU cores and a total of 104.8 TiB of memory;
MPI enables parallel computations across multiple nodes, even if they are of different types.
The login nodes for TOKI-RURI consist of eight servers,
each equipped with dual Intel Xeon Gold 6240 processors, 384 GiB of memory, and one NVIDIA Quadro P4000.
These login nodes are also used for small-scale testing and debugging.
The sequential write to an HDD region in TOKI-FS on a login node is $\simeq 140 \ \textrm{MB}/\textrm{s}$.

\begin{figure}
 \centering
 \includegraphics[clip,keepaspectratio,width=\linewidth]{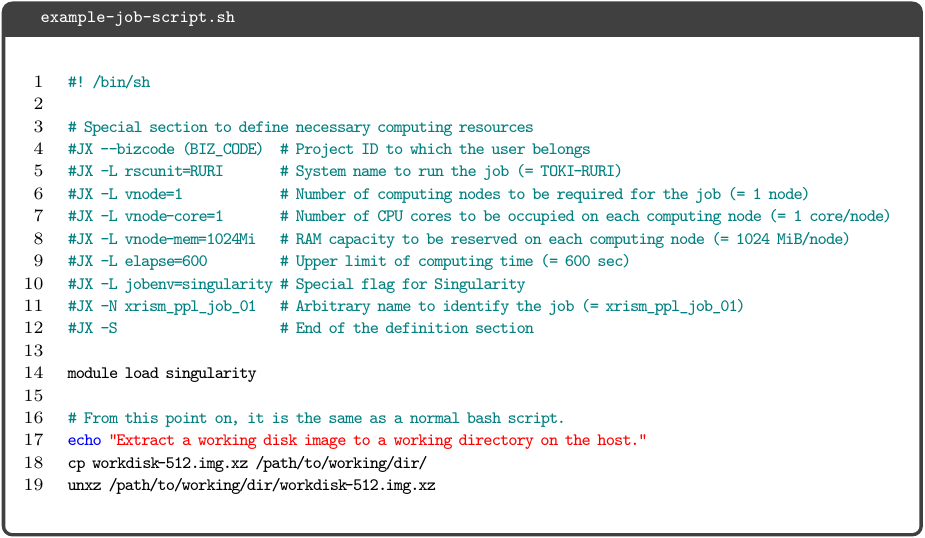}
 \caption{Example of a job script for the JSS3 system.
 Resource requirements for a batch job are provided through comments that start with the special prefix ``\texttt{{\#}JX}.''
 \label{code:jobscript-example}}
\end{figure}

To execute commands (or programs) in parallel on TOKI-RURI, users need to write
and submit a bash script, often referred to as a ``job script,'' that defines necessary computing resources
such as the number of CPU cores, memory capacity, and computation time (Fig.~\ref{code:jobscript-example}).
By considering the resource requirements outlined in the job script and ensuring fairness among users,
a job scheduler allocates the appropriate computing nodes from the four types of TOKI-RURI
listed in Table~\ref{tab:toki-ruri} and determines the execution timing (fair-share scheduling).
This ensures that the job runs without being affected by performance degradation caused by other jobs.
However, the job will be deferred when sufficient resources are not available at that moment.
This implies that the resource requirements of the job script must always exceed the actual usage for the task.
If the actual usage slightly exceeds that defined in the job script, the job is forcibly terminated.
In addition, the user does not know exactly when the job will start
($\simeq 20$ jobs started execution immediately after submitting the job scripts in our case).
Unless an exception request for special circumstances is submitted, the computing time is limited to 24 hours.

Because PPL does not have parallel execution parts, we can reliably request a single CPU core
as a constant value for the job script.
To adapt PPL for TOKI-RURI, we need to derive formulas that estimate the memory capacity and computing time
required for each OBSID based on its observation duration and prepare a metascript to incorporate these values
into a job script template.

\subsection{Virtualization Technologies on Linux and Singularity}

VMs such as the Oracle VirtualBox\cite{VirtualBox} are among the most well-known VTs.
VMs were invented in the late 1960s and are categorized into Type 1 and 2 hypervisors\cite{OracleVMConcept}.
The former runs directly on PC hardware, whereas the latter runs on a host operating system (OS).
Both hypervisors can emulate commonly used PC hardware to run arbitrary OSes.
While PC users can simultaneously execute both Linux and Windows applications, for instance,
on their laptops by employing VMs, running multiple VMs on the respective high-performance servers increases their density
in a data center (Reformatter is the case), thus maximizing the hardware resource usage.
In return, there are overheads due to hardware emulations, which are significant in file input/output (I/O)
to virtual disks.

Another VT isolates a process, an execution unit of programs, and its subprocesses from the others
on an OS.
It originates from the \texttt{chroot} command/system call of Version 7 Unix, released in 1979,
leading to the \texttt{jail} mechanism of FreeBSD introduced in 2000\cite{BSDJails}.
The same feature was implemented as a set of patches, ``Linux-VServer,'' to the Linux kernel, announced in 2001\cite{LinuxVServer}.
Regarding Linux, the process isolation technology known today as ``containers'' is based on the two features
of the Linux kernel \texttt{cgroups}\cite{cgroups} and \texttt{namespaces}\cite{namespaces}.
The former hierarchically groups the respective processes and limits their hardware usage,
such as CPU time, I/O bandwidth, and memory capacity, and the latter constrains their visibilities of software
resources such as mount points, hostname, and user/group IDs.
As the program inside a container instance runs directly on the host kernel, there is marginal overhead,
which is an advantage over VMs.
On the other hand, note that a container on a Linux system accepts only Linux programs, unlike VMs.

Container platforms, such as Docker\cite{Docker} and Singularity
(and its descendant Apptainer\cite{Apptainer}), wrap the \texttt{cgroups} and \texttt{namespaces}
into an easy-to-use form and package all necessary software into a container image
by resolving their mutual dependencies.
Both enable users to build their original container image from a seed image on Docker Hub\cite{DockerHub},
which covers major Linux distributions and creates package management tools such as \texttt{apt}, \texttt{dnf},
\texttt{pacman}, and \texttt{zypper} commands available inside the container instance.
Hence, VMs and containers are similar in terms of practical usage.
Because a Docker instance occupies its image, we can create just one instance from one Docker image;
Docker is frequently used to develop and deploy server applications.
Singularity mounts a container image as a read-only image, which means that multiple instances can be spawned
from a single container image.
Singularity is often used in HPC systems, and its Version 3.10 is also available in TOKI-RURI.

\subsection{Obstacles to the HPC Version of PPL and the Key to the Solution}

In the early design phase of PPL, we fixed all software versions, including the operating system
(RedHat Enterprise Linux 7 with glibc 2.17), compilers (GCC 4.8 and 8),
and scripting languages (Perl 5.16, Python 3.8, and Ruby 2.6).
We also decided that it should only run on Reformatter and is not guaranteed to work in any other environment.
While these compromises in the PPL specifications clarified and simplified the development goals,
the latest version of PPL, its related software, and the auxiliary files have been deployed
in a subdirectory named after the eight-digit release date in the appropriate directory
and identified as ``\texttt{latest}'' by a symbolic link
(note that we mention deployment here, and the source codes are naturally version-controlled by Git).
There were also cross-references between the directories through symbolic links.
Although everything worked perfectly inside Reformatter, these tricks seemed to hinder
the execution of PPL on the TOKI-RURI system because it was time-consuming to manually build all the software
that PPL depends on and to fit PPL into the system's directory structure.

However, based on our knowledge of containers and their technological backgrounds, we provide
an elegant solution for porting PPL to TOKI-RURI:
manipulating the \texttt{mount} \texttt{namespace} and creating everything inside a container instance
identical to Reformatter.
If this is possible, all we have to do is copy the binaries compiled on Reformatter into a container image
and other materials onto TOKI-FS;
Singularity accepts the ``\texttt{--bind}'' option.
For example,
\begin{center}
 \verb|--bind /host/path/to/A:/container/path/to/B|
\end{center}
makes the directory \path{/host/path/to/A} on the host appear as \path{/container/path/to/B}
inside the container.
This option also accepts a loopback device (or disk image) formatted as ext3.
\begin{center}
 \verb|--bind /host/loop/ext3.img:/container/path/to/B:image-src=/|
\end{center}
shows the root directory \path{/} inside the disk image of the host \path{/host/loop/ext3.img}
as a \path{/container/path/to/B} inside the container.

\subsection{Estimating Computing Resources on Reformatter}

During the commissioning period, we needed to profile PPL on Reformatter.
Because no profilers were installed there and we did not have root privileges for Reformatter,
we wrote a small profiler in Python.
The uses and functions of the script are as follows:

We start two terminals and execute the top-level script of PPL in one of them.
Immediately afterward, we run the \texttt{ps} command in the other terminal to find the process ID of the top-level script
and execute the profiler by specifying the process ID as an argument.
The profiler runs the following shell one-liner every minute, where \texttt{pid\_parent} represents the process ID of the top-level script:
Obtain a list of all the descendant processes' process IDs.

\begin{center}
\begin{varwidth}{\linewidth}
\begin{verbatim}
pstree -p (pid_parent) | perl -ne 'print "$1 " while /\((\d+)\)/g'
\end{verbatim}
\end{varwidth}
\end{center}

Next, the profiler immediately executes the following commands to obtain a list of each process's
physical memory usage (Resident Set Size; RSS), reserved memory size (Virtual Memory Size; VSZ), and command name with its arguments.
The variable \texttt{pid\_list} corresponds to a comma-separated list of the aforementioned process IDs.

\begin{center}
\begin{varwidth}{\linewidth}
\begin{verbatim}
ps -p (pid_list) -o pid,rss,vsz,command
\end{verbatim}
\end{varwidth}
\end{center}

The profiler appends these values and the time required for each process to a log file as tab-delimited text
(one line for each process).
The profiler also records the totals of RSS and VSZ at that moment as the consumption of a virtual ``TOTAL'' command
(the name has no other meaning than as a flag) in the log file.
When the top-level script terminates, the profiler loads the log file, groups the data using the command names,
computes the maxima of RSS and VSZ for each command, and stores the results in a summary file.
We consider the maximum VSZ value of the ``TOTAL'' command in the summary file to represent the memory consumption of PPL;
we define the execution time of PPL as the last time recorded in the log file minus the first time.

We selected five OBSIDs that were likely to cover a range of future observation durations and
measured the execution time ($T_{\textrm{Reformatter}}$) and memory usage ($C_{\textrm{Reformatter}}$) of PPL
for each OBSID using the aforementioned method.
Note that Reformatter is noisy regarding these measurements because programs related to XRISM other than PPL are running.
By linear fitting, we obtained
\begin{gather}
 T_{\textrm{Reformatter}} \simeq 0.05041 \, T_{\textrm{Obs}} + 2522 \quad \text{(sec)}, \label{eq:Treformatter} \\
 C_{\textrm{Reformatter}} \simeq 0.05349 \, T_{\textrm{Obs}} + 5104 \quad \text{(MiB)}, \label{eq:Creformatter}
\end{gather}
where $T_{\textrm{Obs}}$ denotes the observation duration in seconds.

C-SODA provides a program in the PPL chain that converts telemetry packets into raw values in a FITS file.
We observed its anomalous behavior in 2020.
An analysis using Valgrind revealed that the initial memory size and increment were not optimized for the typical size of a XRISM RPT file,
wasting approximately half the execution time of the program ($\simeq 400$ s) and $\simeq 4$ GiB of memory.
Although we reported this issue to C-SODA, which was later resolved, the update was not reflected in PPL
because we had already fixed the software version at that time and did not require the absolute performance.
Considering that the program is called multiple times in the PPL chain, the constants in Eqs.~\ref{eq:Treformatter} and \ref{eq:Creformatter}
can be attributed to this overhead.

\subsection{Porting Strategies}

\begin{figure}
 \centering
 \includegraphics[clip,keepaspectratio,width=0.95\linewidth]{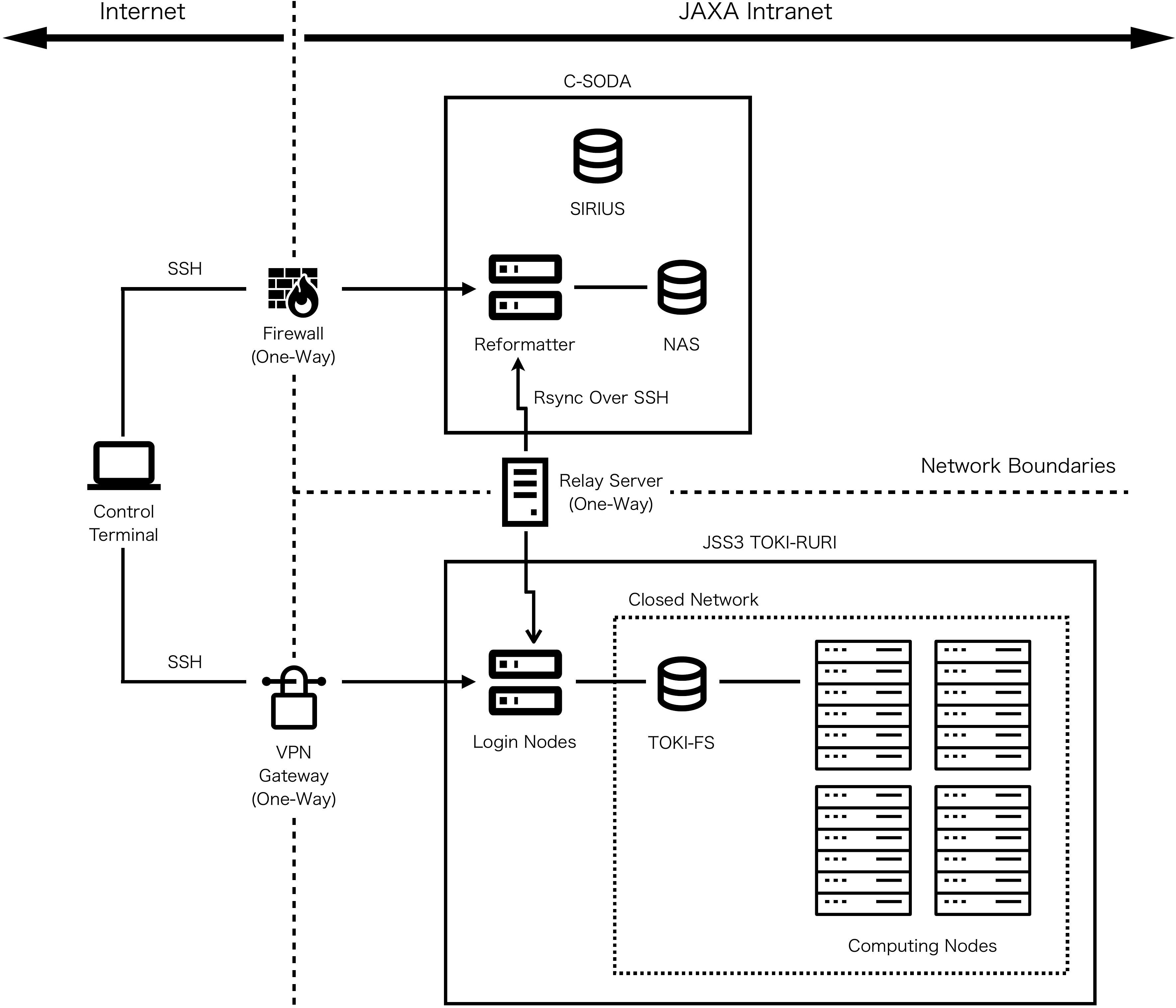}
 \caption{Network overview between Reformatter and TOKI-RURI during HPC PPL processing.
 An operator uses a local PC on the Internet to log in to Reformatter and one of TOKI-RURI’s login nodes through SSH.
 Reformatter and the TOKI-RURI login node are isolated from each other,
 and file transfer between the NAS for Reformatter and TOKI-FS is relayed by the relay server
 over a one-way SSH session from Reformatter to the login node,
 even when the outputs of HPC PPL on TOKI-RURI are rsynced to the NAS
 (the file transfer direction is opposite to that of the SSH session in this case).
 \label{fig:network-overview}}
\end{figure}

\begin{figure}
 \centering
 \includegraphics[clip,keepaspectratio,width=0.89\linewidth]{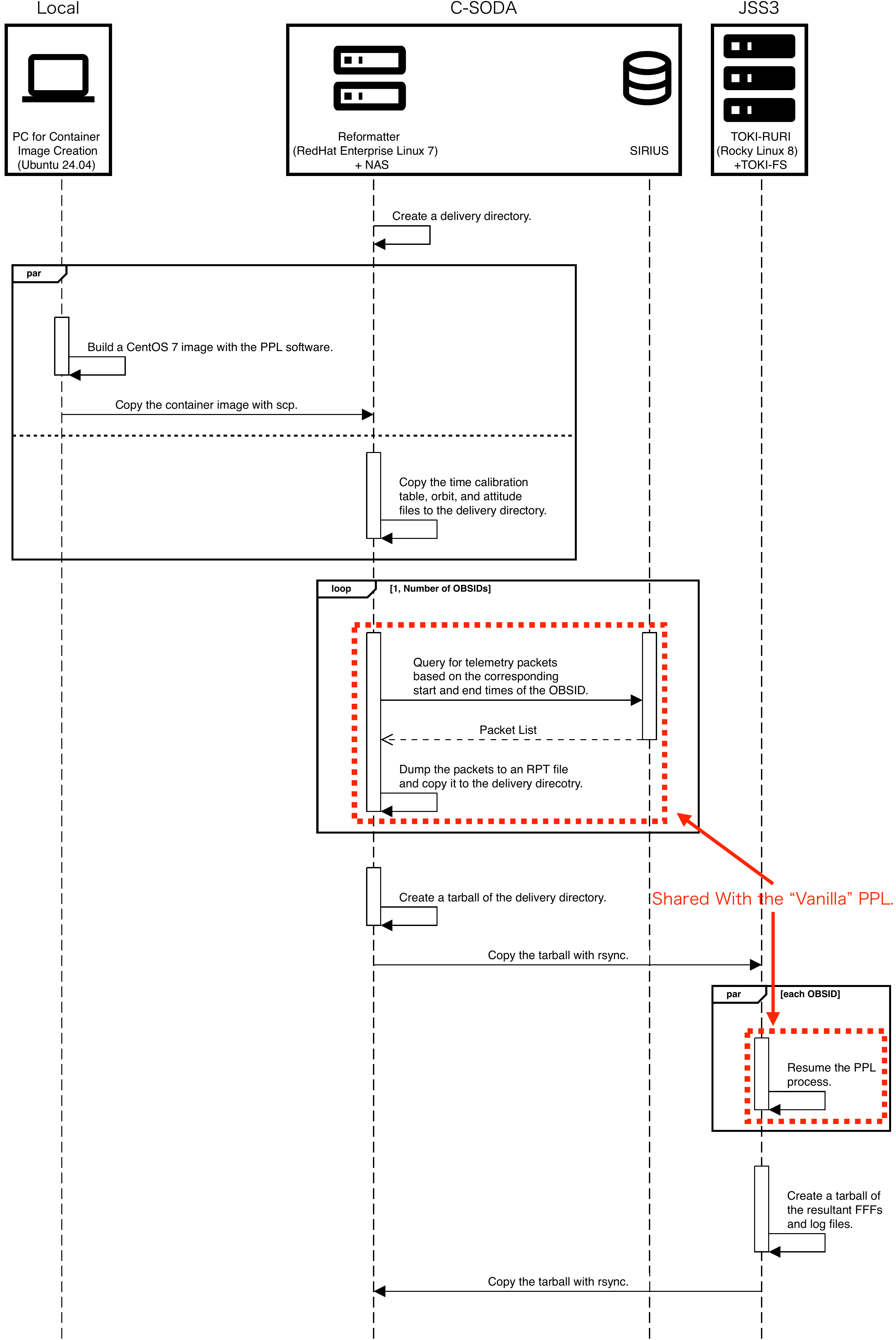}
 \caption{Sequence diagram of the HPC version of pre-pipeline.\label{fig:sequence-of-hpc-ppl}}
\end{figure}

Figures~\ref{fig:network-overview} and \ref{fig:sequence-of-hpc-ppl} illustrate the schematic network setup and sequence diagram of the HPC version of PPL, respectively.
Reformatter has a $\simeq 90 \ \textrm{MB}/\textrm{s}$ (sequential write) Network-Attached Storage (NAS) that stores all input and output data.
We created a directory on the NAS to accumulate all the materials necessary for PPL processing on TOKI-RURI
(the ``delivery directory'' hereafter).
Because TOKI-RURI cannot communicate with SIRIUS, we wrote a small patch (25 lines) to PPL that pauses the PPL processing
when the corresponding RPT file is generated on Reformatter and resumes the remaining processing on TOKI-RURI.

\begin{figure}
 \centering
 \includegraphics[clip,keepaspectratio,width=\linewidth]{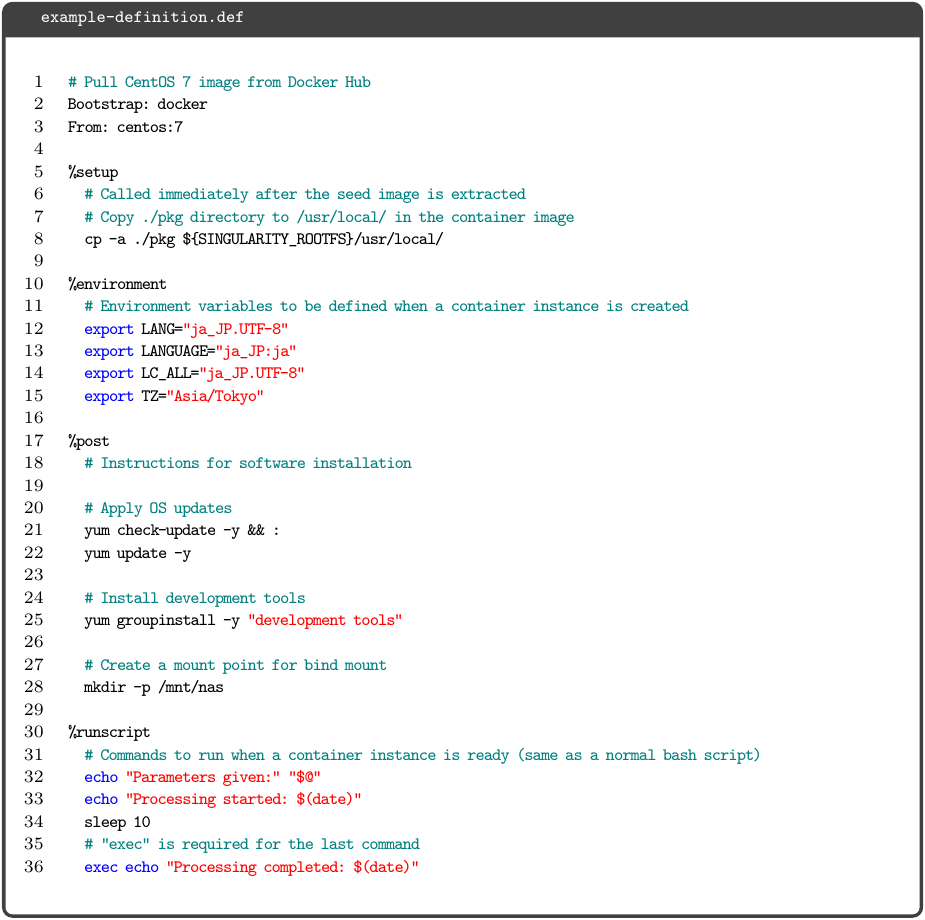}
 \caption{Example of a Singularity definition file.
 A CentOS 7 image is pulled from Docker Hub, all software updates are applied, and development tools are installed.
 When a container instance starts, environment variables for Japanese are set, and \texttt{echo} commands are run.
 \label{code:definitionfile-example}}
\end{figure}

Because we do not have root privileges for Reformatter, we created a container image on a PC other than Reformatter,
where Ubuntu 24.04 is running and Version 3.10.5 of the Singularity Community Edition is installed to match the version with TOKI-RURI.
In the definition file, a recipe for creating the container image, a seed of CentOS 7 on Docker Hub, is obtained,
and all software updates are applied;
all software packages installed on Reformatter with the \texttt{yum} command are also installed similarly on the container.
At the end of the definition file, the binaries of the patched PPL built on Reformatter are copied to the container image,
and all mount points for the ``\texttt{--bind}'' option are created (Fig.~\ref{code:definitionfile-example}).
The image ($\simeq 7$ GiB) is transferred to the delivery directory on Reformatter via the \texttt{scp} command.

On Reformatter, the time calibration table, orbit, and attitude files are copied to the delivery directory.
A small Python script reads the list of OBSIDs, sequentially creates the respective RPT files,
and copies them to the delivery directory.
The directory is archived into a tarball with Zstandard compression ($\sim$ TB) and copied to a TOKI-RURI login node using the \texttt{rsync} command, allowing file transfer to resume in case of unexpected network interruptions.
The bandwidth of \texttt{rsync} was approximately $\simeq 50$ MiB/s.

When the file transfer is completed, it is manually extracted into the working directory of TOKI-FS (not SSDs but HDDs).
As described in Sec.~\ref{sect:jss3}, we require a job script that defines the computing resources
necessary to run the containerized PPL as a batch job on TOKI-RURI.
The key point here is not that the estimation is correct but rather that the actual resource consumption does not exceed the estimation.
By considering the difference in the CPU clock frequencies between Reformatter and TOKI-RURI,
the overhead of the Lustre FS (Sec.~\ref{sect:v2-improvements}),
and margins, we estimate the computing time $T_{\textrm{TOKI}}$ and memory requirement $C_{\textrm{TOKI}}$ on TOKI-RURI as follows:
\begin{gather}
 T_{\textrm{TOKI}} = 4 \, T_{\textrm{Reformatter}} \  , \label{eq:Ttoki} \\
 C_{\textrm{TOKI}} = 2 \, C_{\textrm{Reformatter}} \  ,
\end{gather}
with the constaints $8 \le T_{\textrm{TOKI}} \le 120$ (hours) and $32 \le C_{\textrm{TOKI}} \le 176$ (GiB).
If we simply apply Eq.~\ref{eq:Ttoki}, $T_{\textrm{TOKI}}$ exceeds 24 hours,
which is the default maximum execution time.
Therefore, we applied for an extension to this limit.

A Python script reads the list of OBSIDs with their observation start and end times, computes the respective
$\left( T_{\textrm{TOKI}}, C_{\textrm{TOKI}} \right)$,
builds the argument for the ``\texttt{--bind}'' option, which gives the mappings between the native FS and
inside the container on TOKI-RURI, and writes them into the corresponding job scripts.
The Python script automatically submits these job scripts.
When all jobs are completed, the resultant FFFs and log files are manually archived into a tarball using Zstandard compression.
The tarball is sent back to Reformatter with the \texttt{rsync} command and extracted into a specified directory on Reformatter.

\section{Results and Discussion} \label{sect:results}

\begin{table}
 \caption{Number of OBSIDs for reprocessing and the statistics.} \label{tab:statistics}
 \centering
 \begin{tabular}{cccccc}
  \hline
   & HPC PPL & OBSIDs & Reformatter & TOKI-RURI & Failed \\
  Project Period & Version & & (hours) & (hours) & OBSIDs \\
   & (1) & (2) & (3) & (4) & (5) \\
  \hline
  Commissioning & 1 & 80 & 218.3 & 26.3 & 11 \\
  Comissioning$+$PV & 2 & 161 & 515.5 & 15.2$^{\dagger}$ & 0 \\
  \hline
  \multicolumn{5}{l}{{\small {${}^{\dagger}$} Disk images formatted to ext3 were used for working areas.}} \\
  \multicolumn{5}{l}{{\small (1) The version number of HPC PPL.}} \\
  \multicolumn{5}{l}{{\small (2) The number of OBSIDs for reprocessing.}} \\
  \multicolumn{5}{l}{{\small (3) Estimated total computing time in units of hours on Reformatter.}} \\
  \multicolumn{5}{l}{{\small (4) Actual total computing time in units of hours on TOKI-RURI.}} \\
  \multicolumn{5}{l}{{\small (5) The number of OBSIDs that failed to be reprocessed.}}
 \end{tabular}
\end{table}

\subsection{Reprocessing with HPC PPL Version 1}

In March 2024, we reprocessed the 80 OBSIDs observed during the commissioning period using the initial version of the HPC PPL.
The results are summarized in the first row of Table~\ref{tab:statistics}.
It was estimated to take 218.3 hours to complete the reprocessing tasks on the Reformatter with no parallelization;
the time required to create the RPT files was estimated to be $\simeq 3\%$ of the total.
The HPC PPL completed them in 26.3 hours (the latest log file update time among all 80 jobs
minus the earliest creation time), although it failed to process 11 OBSIDs (Sec.~\ref{sect:v1-failures} for the cause),
which were later reprocessed on Reformatter.
When we define speedup $S$ as
\begin{equation}
 S = \dfrac{\text{(Estimated Total Computing Time on Reformatter)}}{\text{(Actual Total Computing Time on TOKI-RURI)}} \ ,
\end{equation}
and subtract the time for the RPT file creation from the estimated computing time on Reformatter
(as the HPC PPL does not create them), we obtain $S = 218.3 \cdot \left( 1 - 0.03 \right) / 26.3 = 8.1$;
two PPL tasks can run simultaneously on Reformatter, and $S$ is reduced to $S = 4.0$ in this case;
however, this is a decent speedup.

\subsection{Cause of the Failures} \label{sect:v1-failures}

Shortly after completing the reprocessing tasks, we investigated the causes of the failures in detail.
We successfully reproduced these errors on a PC for container image creation.
To convert the CCSDS packets in an RPT file into raw values, we used a set of programs provided by C-SODA.
For some of them, no source code was accessible for SOC, and some of these without the source code were 32-bit binaries.
One of these 32-bit programs invokes the others and communicates with them through Standard Input/Output.
Because these 32-bit programs are unaware of the 64-bit inode numbers, which is the case for XFS and Lustre FS,
they fail to read their configuration files, open the input and output files, and spawn others onto these FSes
when their inode numbers exceed the 32-bit range.
One of the advantages of containers is that we can perform detailed analyses on our own PCs
in which various debugging tools are installed.

\subsection{Improvements Toward Version 2} \label{sect:v2-improvements}

Our solution to this problem is straightforward: we used disk images formatted to ext3, where everything was 32-bit,
as working areas.
We attached two working disk images to each container instance with the ``\texttt{--bind}'' option:
one 64 GiB for the PPL software and another 512 GiB for the RPT and FFFs because the option does not accept the mapping
from a disk image on the host onto two different directories inside the container instance.
Although copying the PPL software inside the container image onto an ext3 disk image for each container instance
would be wasteful, this reduces human errors during container image creation.
Because PPL is required to process continuous observations for up to 10 days and we have never experienced such a condition,
we could not determine whether 512 GiB is excessive.

Using a disk image as the working area is also expected to improve the I/O performance of the Lustre FS,
which consists of MetaData Servers (MDSes) and Object Storage Servers (OSSes).
When an append operation occurs in an existing file, the MDSes search for adequate blocks among the storage devices mounted by OSSes
and update the metadata shared among the MDSes;
when small appends occur extensively in a single file, which is the case for log and FITS file manipulations,
the performance of Lustre FS is significantly degraded\cite{Lee2023}.
However, when partial changes occur within a file for which space has already been allocated, the computation for block allocations
shall be omitted, leading to minimal performance degradation.

The \texttt{dd} command can create a disk image with general user permission, formatting the image to ext3
with the \texttt{mkfs.ext3} command requires root permission.
Hence, we format the images on the local PC and compress them with the \texttt{xz} command with the ``\texttt{-9}'' option;
this reduces the 64 and 512 GiB images to 10 and 77 MiB, respectively.
The compressed working disks and container images are transferred to the delivery directory
on the NAS attached to Reformatter.
When a container instance is invoked on TOKI-RURI, the PPL software inside the container image and the RPT file
on the host are copied to their respective ext3 working disks.
After the PPL process is completed and immediately before the container instance is terminated,
the FFFs as products and log files are synchronized with the host FS.
Considering the time required for synchronization,
Eq.~\ref{eq:Ttoki} is modified as
\begin{equation}
 T_{\textrm{TOKI}} = 4 \, T_{\textrm{Reformatter}} + 4 \quad \text{(hours)}, \label{eq:Ttoki_v2}
\end{equation}
with constraint $12 \le T_{\textrm{TOKI}} \le 24$ (h).
While we cannot benchmark the TOKI-RURI environment due to the scheduler described in Sec.~\ref{sect:jss3},
testing on a local machine indicated that these bootstrapping
and shutdown processes resulted in a 25-minute increase over the 100-minute main processing time.
Even considering this increased execution time, our understanding of the bottleneck in Lustre FS indicates
that Eq.~\ref{eq:Ttoki_v2} is an overestimate;
the actual computing time should be within 24 hours, which is the default maximum computing time for TOKI-RURI
(four hours as insurance in case our prediction is incorrect).

\subsection{Reprocessing with HPC PPL Version 2}

In September 2024, soon after the PV period, we reprocessed the 161 OBSIDs observed
during the commissioning and PV periods, respectively, using the second version of the HPC PPL.
Because running 161 jobs at once could reach the disk quota of the working directory on TOKI-RURI,
we submitted them in two batches: 100$+$61;
the latter batch was submitted when 90\% of the former was completed.
The results are summarized in the second row of Table~\ref{tab:statistics}.
While it was estimated to take 515.5 hours to complete the reprocessing tasks on Reformatter
with no parallelization, the HPC PPL completed them within 15.2 hours;
it is noteworthy that the operator was not constantly monitoring the execution status.
The speedup $S$ is
\begin{equation}
 S = \dfrac{515.5 \cdot \left( 1 - 0.03 \right)}{15.2} = 33.
\end{equation}
No errors were observed.
From this, the following conclusions were drawn:
\begin{itemize}
 \item We successfully ported the PPL to the HPC system using Singularity and its ``\texttt{--bind}'' option.
 \item A $33 \times$ speedup at least was accomplished;
       three-week jobs were reduced to about half a day.
 \item Using a working disk image formatted to ext3 maximizes the Lustre FS’s performance
       even for software designed for a standard Linux system, not intended for an HPC system;
       Eq.~\ref{eq:Ttoki} can finally be reduced to $T_{\textrm{TOKI}} \simeq T_{\textrm{Reformatter}}$.
\end{itemize}

\subsection{Bottleneck: Bandwidth of the File Transfers}

We mentioned file transfers during reprocessing in September.
The size of the tarball included all RPT files created on the Reformatter,
as well as the Singularity and working disk images, was 1.6 TiB.
The size of the tarball containing all the products on TOKI-RURI was 1.2 TiB.
When the bandwidth of 50 MiB/s is assumed, we obtain 9.3 and 7.0 hours for the file transfers, respectively.
Even if these are added to the computing time of TOKI-RURI, we could still conclude that the three-week jobs were reduced to two days.
On the other hand, a file transfer with parallel TCP streams using the \texttt{bbcp}\cite{bbcp} command will shorten this time.
However, these connections are now completely blocked by the firewall on the JSS3 side.
Reducing the file transfer time will require coordination between departments within JAXA,
which is expected to be realized in the future.

\section{Summary} \label{sect:summary}

We briefly described the regular data processing in XRISM, which consists of PPL and PL.
As both software programs are still being improved and scientists cannot apply the latest versions of PPL and CALDB
to their data owing to the non-publicity of FFFs, SOC has to reprocess all the observation data
for each transition during the study period.
We developed porting methods to boost the number of reprocessing tasks that enable PPL to run on the TOKI-RURI HPC system
using Singularity, which is a container platform.
Using these methods, PPL, which only supports RedHat Enterprise Linux 7,
can operate on Rocky Linux 8 (TOKI-RURI) and Ubuntu 24.04 (the PC used to create the container image).
Using disk images formatted to ext3 as working areas, we obtained a $33 \times$ speedup even on the Lustre FS.
File transfer over the Internet between Reformatter and TOKI-RUI is currently a bottleneck;
however, parallel TCP connections can be used as a solution.

The total amount of observational data is proportional to the duration of the mission.
Although we anticipate that XRISM will continue its observations for over three years,
no mission schedule has been finalized beyond September 2026.
Therefore, there should be $\simeq 160 \times 3 = 480$ OBSIDs,
and the reprocessing time on Reformatter is estimated to exceed two months;
our methods will reduce it to one week.
Furthermore, the proposed methods are highly versatile;
based on the methods, it will be ordinal for observational data obtained with a scientific satellite
to be processed with an HPC system.

\section*{Code and Data Availability}

The source codes of PPL (including its HPC version) and PL are proprietary and not publicly available.
The data supporting this article's findings are also publicly unavailable because the PPL software and
its running environment are required.
The subset of data underlying Eqs.~\ref{eq:Treformatter} and \ref{eq:Creformatter} and
Table~\ref{tab:statistics} can be requested from the author at \linkable{sa-eguchi@kumagaku.ac.jp}.

\section*{Disclosures}

The authors declare no financial or commercial conflicts of interest,
or other potential conflicts of interest that could have influenced the objectivity
of this study and the writing of this paper.

\section*{Acknowledgments}

This work was supported by the Japan Society for the Promotion of Science (JSPS) Core-to-Core Program (Grant Number: JPJSCCA20220002)
and the Japan Society for the Promotion of Science Grants-in-Aid for Scientific Research (KAKENHI)
(Grant Numbers JP20K04009 (YT), JP21K03615, and JP24K00677 (MN)).
The work of ML and TY was supported by NASA under the award number 80GSFC24M0006.

Grammarly was used to improve language and grammar.
We would like to thank Editage (\texttt{www.editage.jp}) for English language editing.

%%%%% References %%%%%

\bibliography{report}   % bibliography data in report.bib

\begin{thebibliography}{10}

\bibitem{Audard2024}
M.~{Audard}, H.~{Awaki}, R.~{Ballhausen}, {\em et~al.}, ``{The XRISM
  first-light observation: Velocity structure and thermal properties of the
  supernova remnant N 132D},'' {\em \pasj}   (2024).

\bibitem{xrism_ngc4151}
{XRISM Collaboration}, M.~{Audard}, H.~{Awaki}, {\em et~al.}, ``{XRISM
  Spectroscopy of the Fe K{\ensuremath{\alpha}} Emission Line in the Seyfert
  Active Galactic Nucleus NGC 4151 Reveals the Disk, Broad-line Region, and
  Torus},'' {\em \apjl} {\bf 973}, L25  (2024).

\bibitem{Mori2024}
K.~{Mori}, H.~{Tomida}, H.~{Nakajima}, {\em et~al.}, ``{Status of Xtend
  telescope onboard X-Ray Imaging and Spectroscopy Mission (XRISM)},'' in {\em
  Space Telescopes and Instrumentation 2024: Ultraviolet to Gamma Ray},
  J.-W.~A. {den Herder}, S.~{Nikzad}, and K.~{Nakazawa}, Eds., {\em Society of
  Photo-Optical Instrumentation Engineers (SPIE) Conference Series} {\bf
  13093}, 130931I  (2024).

\bibitem{Terada2021}
Y.~{Terada}, M.~{Holland}, M.~{Loewenstein}, {\em et~al.}, ``{Detailed design
  of the science operations for the XRISM mission},'' {\em Journal of
  Astronomical Telescopes, Instruments, and Systems} {\bf 7}, 037001  (2021).

\bibitem{Kurtzer2017}
G.~M. Kurtzer, V.~V. Sochat, and M.~W. Bauer, ``Singularity: Scientific
  containers for mobility of compute,'' {\em PLoS ONE} {\bf 12}  (2017).

\bibitem{JSS3}
{SuperComputer Division, Japan Aerospace Exploration Agency}, ``{Overview of
  JSS3}.'' {\url{https://www.jss.jaxa.jp/en/}}  (2024).
\newblock (Accessed on 11/1/2024).

\bibitem{Eguchi2022}
S.~{Eguchi}, M.~{Tashiro}, Y.~{Terada}, {\em et~al.}, ``{Xappl: software
  framework for the XRISM pre-pipeline},'' in {\em Space Telescopes and
  Instrumentation 2022: Ultraviolet to Gamma Ray},  J.-W.~A. {den Herder},
  S.~{Nikzad}, and K.~{Nakazawa}, Eds., {\em Society of Photo-Optical
  Instrumentation Engineers (SPIE) Conference Series} {\bf 12181}, 1218161
  (2022).

\bibitem{Yamada2008}
T.~Yamada, ``Standardization of spacecraft and ground systems based on a
  spacecraft functional model,'' in {\em SpaceOps 2008 Conference},   (2008).

\bibitem{Okada2012}
N.~Okada and Y.~Yamamoto, ``Concept for next-generation science satellite's
  telemetry database,'' {\em JAXA Research and Development Report: Journal of
  Space Science Informatics Japan} {\bf 1}, 151  (2012).
\newblock {(in Japanese)}.

\bibitem{Wells1981}
D.~C. {Wells}, E.~W. {Greisen}, and R.~H. {Harten}, ``{FITS - a Flexible Image
  Transport System},'' {\em \aaps} {\bf 44}, 363  (1981).

\bibitem{Pence2010}
W.~D. {Pence}, L.~{Chiappetti}, C.~G. {Page}, {\em et~al.}, ``{Definition of
  the Flexible Image Transport System (FITS), version 3.0},'' {\em \aap} {\bf
  524}, A42  (2010).

\bibitem{Matsuzaki2012}
K.~MATSUZAKI, T.~SAITO, S.~OKUNISHI, {\em et~al.}, ``Automatic generation of
  on-board software from the model - spacecraft information base version 2
  (sib2),'' {\em TRANSACTIONS OF THE JAPAN SOCIETY FOR AERONAUTICAL AND SPACE
  SCIENCES, AEROSPACE TECHNOLOGY JAPAN} {\bf 10}(ists28), Tf\_11--Tf\_17
  (2012).

\bibitem{A64FX}
{Fujitsu Limited}, ``{A64FX Microarchitecture Manual}.''
  {\url{https://github.com/fujitsu/A64FX/blob/master/doc/A64FX_Microarchitecture_Manual_en_1.3.pdf}}
   (2020).

\bibitem{Lustre}
{Open Scalable File Systems, Inc.}, ``{Lustre File System}.''
  {\url{https://www.opensfs.org/lustre/}}  (2024).
\newblock (Accessed on 11/20/2024).

\bibitem{VirtualBox}
{Oracle}, ``{Oracle VirtualBox}.'' {\url{https://www.virtualbox.org/}}  (2024).
\newblock (Accessed on 11/25/2024).

\bibitem{OracleVMConcept}
{Oracle}, ``{Oracle VM Concepts Guide for Release 3.4}.''
  {\url{https://docs.oracle.com/en/virtualization/oracle-vm/3.4/concepts/index.html}}
   (2021).

\bibitem{BSDJails}
P.-H. {Kamp} and R.~N.~M. {Watson}, ``{Jails: Confining the omnipotent root}.''
  {\url{https://docs-archive.freebsd.org/44doc/papers/jail/jail.html}}  (2000).
\newblock (Accessed on 11/20/2024).

\bibitem{LinuxVServer}
J.~{Gelinas}, ``{Announce: many virtual servers on a single box}.''
  {\url{https://lkml.iu.edu/hypermail/linux/kernel/0110.1/0909.html}}  (2001).
\newblock (Accessed on 11/21/2024).

\bibitem{cgroups}
J.~{Corbet}, ``{Process containers}.'' {\url{https://lwn.net/Articles/236038/}}
   (2007).
\newblock (Accessed on 11/21/2024).

\bibitem{namespaces}
M.~{Kerrisk}, ``{Namespaces in operation, part 1: namespaces overview}.''
  {\url{https://lwn.net/Articles/531114/}}  (2013).
\newblock (Accessed on 11/21/2024).

\bibitem{Docker}
{Docker Inc.}, ``{Docker: Accelerated Container Application Development}.''
  {\url{https://www.docker.com/}}  (2024).
\newblock (Accessed on 11/25/2024).

\bibitem{Apptainer}
{Apptainer Container Project}, ``{Apptainer - Portable, Reproducible
  Containers}.'' {\url{https://apptainer.org/}}.
\newblock (Accessed on 11/25/2024).

\bibitem{DockerHub}
{Docker Inc.}, ``{Docker Hub Container Image Library}.''
  {\url{https://hub.docker.com/}}.
\newblock (Accessed on 11/25/2024).

\bibitem{Lee2023}
C.~Lee, J.~Lee, C.~kim, {\em et~al.}, ``I/o separation scheme on lustre
  metadata server based on multi-stream ssd,'' {\em Cluster Computing} {\bf
  26}, 2883--2896  (2023).

\bibitem{bbcp}
{Board of Trustees of the Leland Stanford, Jr. University}, ``{BBCP}.''
  {\url{https://www.slac.stanford.edu/~abh/bbcp/}}  (2014).
\newblock (Accessed on 12/3/2024).

\end{thebibliography}
\bibliographystyle{spiejour}   % makes bibtex use spiejour.bst

%%%%% Biographies of authors %%%%%

\vspace{2ex}\noindent\textbf{Satoshi Eguchi} is an associate professor in the Department
of Economics at Kumamoto Gakuen University.
He received his BS and MS degrees and PhD in science from Kyoto University in 2006, 2008,
and 2011, respectively.
His areas of interest include active galactic nuclei and software development for data analysis
(or computing) in astronomy, including in virtual observatories.
He developed ``ALMAWebQL'' and ``SNEGRAF,''  web applications for radio and
gravitational-wave astronomy, respectively.

\vspace{1ex}
\noindent Biographies and photographs of the other authors are not available.

%%\listoffigures
%%\listoftables

\end{spacing}
\end{document}